\documentclass[conference]{IEEEtran}
\usepackage{cite}
\usepackage{amsmath,amssymb,amsfonts}
\usepackage{algorithmic}
\usepackage{graphicx}
\usepackage{hyperref}
\usepackage{textcomp}
\usepackage{import}
\usepackage{xcolor}
\def\BibTeX{{\rm B\kern-.05em{\sc i\kern-.025em b}\kern-.08em
    T\kern-.1667em\lower.7ex\hbox{E}\kern-.125emX}}

\begin{document}

\title{LightOn Optical Processing Unit: Scaling-up AI and HPC with a Non von Neumann co-processor   \vspace*{-1mm} }

\author{Charles Brossollet, Alessandro Cappelli, Igor Carron, Charidimos Chaintoutis, Am\'elie Chatelain, \\ 
Laurent Daudet, Sylvain Gigan, 	Daniel Hesslow, Florent Krzakala, Julien Launay, Safa Mokaadi, \\ 
Fabien Moreau, 	Kilian Müller,  Ruben Ohana,	Gustave Pariente,
	Iacopo Poli, and Elena Tommasone \\
	{\textit{LightOn, Paris, France }} \\%hskip 1cm
	{\tt contact@lighton.io}
	}

\maketitle

\begin{abstract}
We introduce LightOn's Optical Processing Unit (OPU), the first photonic AI accelerator chip available on the market for at-scale Non von Neumann computations, reaching 1500 TeraOPS. It relies on a combination of free-space optics with off-the-shelf components, together with a software API allowing a seamless integration within Python-based processing pipelines. We discuss a variety of use cases and hybrid network architectures, with the OPU used in combination of CPU/GPU, and draw a pathway towards ``optical advantage''.
\end{abstract}

\section{Introduction}
In recent years, a number of photonic chips for AI computations have emerged \cite{hughes2018,guo2019, ramey2020}, taking advantage of high bandwidth, high parallelism and low energy consumption.
Some of the most advanced designs are based on integrated photonics, typically implementing generic matrix-vector multiplications at GHz rates. These approaches are well suited to applications such as convolutional neural networks for edge computing, but are intrinsically limited to small dimensional signals. 

Here, we take a different approach, and target heavy data-center computations involving extremely high-dimensional signals - up to 1 million. These data appear in many modern Machine Learning applications, such as Graph Neural Networks, Natural Language Processing - based on "transformers" such as GPT-3 -, or neural view synthesis. At these sizes, the "von Neumann bottleneck" becomes more acute, as matrix sizes may outsize the RAM limits, especially in GPUs. 
We here introduce LightOn Appliance, released March 7th, 2021, based on the Optical Processing Unit (OPU) technology.

\section{LightOn's Optical Processing Unit}

The OPU leverages light scattering  \cite{saade2016} to perform, in the analog domain, Random Projections, i.e. the multiplication of input vectors $\bf{x}$ by a {\it fixed} random matrix $M$, whose entries follow an independent and identically distributed complex Gaussian distribution. The output is ${\bf y} = | M \bf{x} |^2$, with element-wise non-linearity $| . |^2$. The built-in non-linearity can also be suppressed by interferometric measurements, leading to ${\bf y} = M \bf{x}$.
The benefits of the OPU comes from the dimensionality of the data, the speed at which these computations are made, and the low power consumption. In the LightOn Appliance OPU, $\bf{x}$ (binary) and $\bf{y}$ (8-bit) scale up to dimension 1 million and 2 million, respectively, and independent computations can be made at 1.9 kHz, for a power consumption of 30 W.
It thus reaches 1500 TeraOPS, or 50 TeraOPS / W.

\begin{figure}%[h]
 \centering
 \vspace*{-6mm}
	\includegraphics[width=80mm]{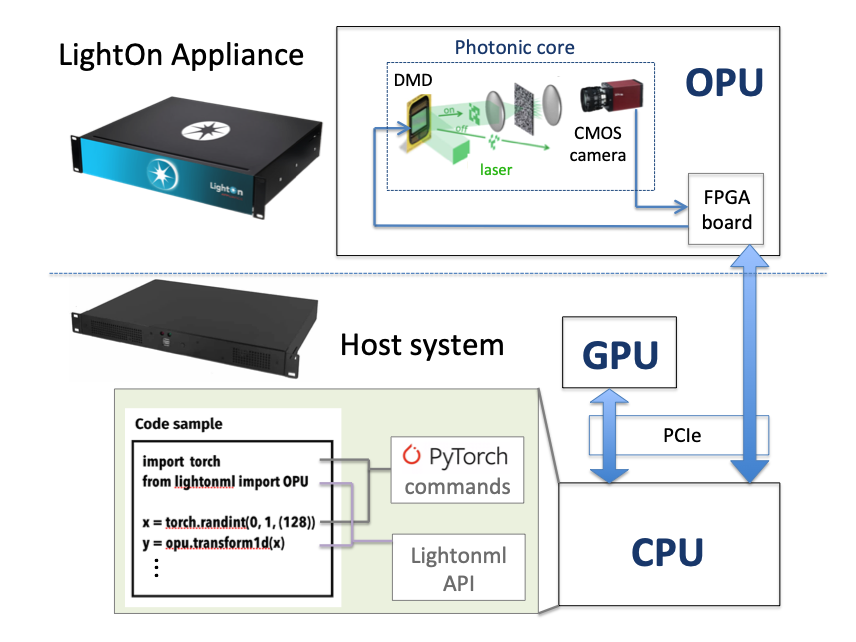} % Figure image
	 \vspace*{-3mm}
	\caption{The hybrid data processing architecture, featuring LightOn's OPU as an external co-processor.} % Figure caption
	\label{system} % Label for referencing with \ref{bear}
	 \vspace*{-3mm}
\end{figure}

The OPU operates in a "Non von Neumann" %(NvN) 
regime: although the $2.10^{12}$ weights of the matrix $M$ are fixed \textit{by design}, they are accessed instantly, at no energy cost: $M$ plays the role of a large read-only memory (terabytes equivalent),  that  can  be  used in matrix multiplications, literally at the speed of light and in a passive way. Speed limitations and power consumption arise as a result of communication  and  formatting, D/A and A/D  conversion, and laser power. In contrast to von Neumann architectures, where computing time and memory requirements scale with the size $n$ of the data, i.e. $O(n^2)$ for a matrix-vector multiplication, the computation time is here $O(1)$ independent on the data size.  
At large $n$ - typically above $10^5$ -, this NvN operation gets faster - but more importantly allows direct single-chip implementation on larger signals without reaching RAM limits.

\paragraph*{Hardware}
The LightOn Appliance OPU is packaged as a 2U rackable device, linked to its host server through Gen2 x4 external PCIe, as shown on Fig. \ref{system}. It contains a single compact photonic core, custom FPGA boards for data i/o, a laser and power supply. All components, including light modulators and detectors, are  mass produced for consumer markets.

\paragraph*{Software}
The software layer has been designed to offer a smooth experience to Machine Learning experts, without any knowledge in photonics. The custom API library {\tt LightOnML}, integrated in Python,  provides pre-processing functions for different types of input data.
This API is compatible with Pytorch and Scikit-learn.

\section{Hybrid Computing Architectures}
\begin{figure}
 \centering
 \vspace*{-1mm}
	\includegraphics[width=80mm]{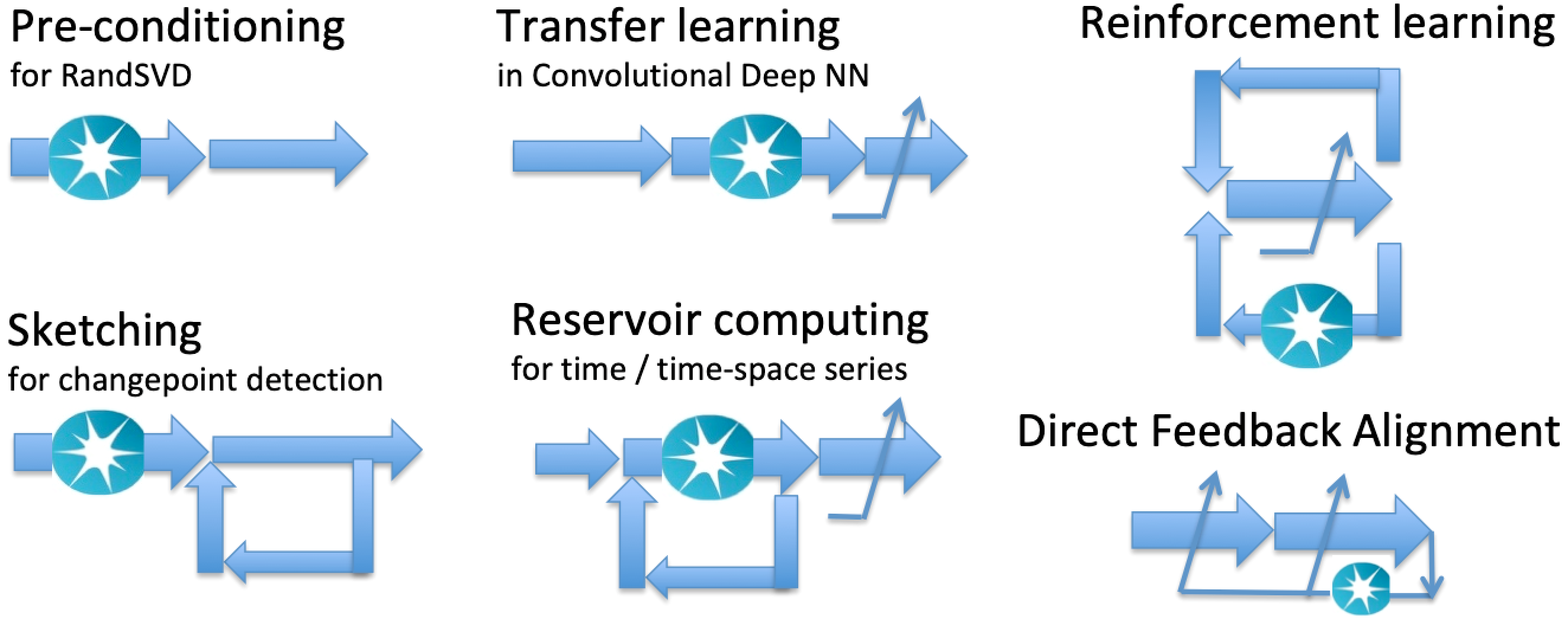} % Figure image
	\caption{Different Neural Network architectures taking advantage of LightOn's OPU - position indicated by the "flare" logo in the hybrid processing pipeline. Other arrows indicate computations performed by CPU or GPU.} % Figure caption
	\label{archi} % Label for referencing with \ref{bear}
	 \vspace*{-3mm}
\end{figure}

\paragraph*{ML applications} Fig. \ref{archi} displays some neural network architectures that use the OPU in hybrid computing pipelines, such as for Natural Language Processing, change-point detection in multi-dimensional time series \cite{keriven2018}, molecular dynamics \cite{chatelain2020}, event classification in particle physics, graph neural networks \cite{hashem2020}  as well as more fundamental studies: supervised random projections or kernel computations\cite{ohana2019kernel}. 
Interestingly, some properties are due to the analog nature of the OPU, such as increased robustness from adversarial attacks  \cite{cappelli2020}.
More details can be found on LightOn's blog \cite{LightOnBlogWWW}, and public GitHub source code repository \cite{LightOnGithub}. As an example of typical speedup, in a Transfer Learning experiment, using the OPU for a dense layer between convolutional features and ridge regression leads to $\times 8$ speedups and $\times 11$ energy savings compared to the same code on CPU/GPU only, with the same final accuracy. This example \cite{LightOndocsTL} can be run directly on the LightOn Cloud.
Finally, let us emphasize the particular case of Direct Feedback Aligment \cite{launay2020}, where the OPU random projections are used in the feedback loop, as an alternative to back-propagation training. This represents, to our knowledge, the only optical training applied to large-scale ($>$ 1 million parameters) modern Neural Network architectures, including Graph Neural Networks \cite{launay2020_NeurIPSworkshop}, or transformers. 

\begin{figure}
 \centering
 \vspace*{-2mm}
	\includegraphics[width=85mm]{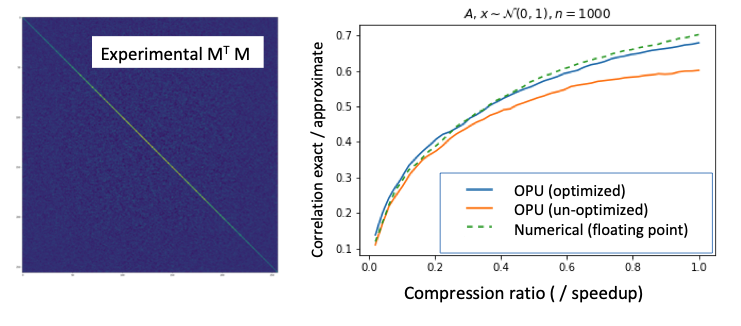} % Figure image
 \vspace*{-2mm}
 \caption{Approximate matrix-vector multiplications (from \cite{hesslow21}). Left: experimental verification of $M^T M \approx I$. Right: approximation vs. compression ratio, comparison of baseline numerical approximation with different OPU schemes} % Figure caption
	\label{amm} % Label for referencing with \ref{bear}
	\vspace*{-2mm}
\end{figure}

\paragraph*{HPC applications: Accelerated Linear Algebra}
Randomized Numerical Linear Algebra is a widely studied technique, to speed-up large computations in various HPC applications such as inverse problems or finance. 
Here, we only discuss how the OPU technology offers an alternative view, and refer to the companion study \cite{hesslow21} for details. 
At the simplest level, for a large random matrix $M$, one has $M^T M \approx I$ (up to normalization). A matrix-vector product $A x$ can be approximated in the compressed domain:  
$A x \approx A ( M^T M ) x = ( M A )^T (M  x)$, assuming that $M$ is fat $m \times n$, with $m < n$. With the OPU, the products $\tilde{A}  = M A$ (pre-computed once, assuming $A$ is fixed) and $\tilde{x} = M x$ can be performed efficiently. Finally, one is left with computing $\tilde{A} \tilde{x}$ in the compressed domain. At sizes where the OPU random projection takes negligible time, approximate matrix-vector multiplication is performed with a speedup $n/m$. 
Fig. \ref{amm} shows that optimized OPU pipelines provide approximate results close to full precision randomization. The same principle has been applied to Randomized SVD \cite{LightOndocsSVD}, that can serve as a basis for recommender systems. For large {\it dense} matrices, such methods may represent the only practical alternative.  

\section{Conclusion: towards ``optical advantage''}
In many ML / HPC computing tasks, not all coefficients need to be updated. Free space photonics is currently the most promising way to leverage the Non von Neumann principle at scale, with instantaneous and energy-passive access to trillion size coefficient arrays.
 With LightOn's OPU, this technology is now mature, seamlessly integrated in standard computing pipelines - as a complement to standard CPU / GPU programmable chips.
Here, we have demonstrated a few examples of hybrid computing. As data and models become larger and larger, the benefit of such technologies becomes clearer: we believe that, in order to scale up already massive language models such as GPT-3, it offers a unique pathway to ``optical advantage'' - i.e. the use of a "beyond pure silicon" technology in business-relevant computations, that would otherwise require dedicated supercomputers.

%\section*{References}
\bibliography{references} 
\bibliographystyle{ieeetr}

\end{document}